\documentclass[10pt, conference, compsocconf]{IEEEtran}
\ifCLASSINFOpdf
\else
\fi

\usepackage{balance}
\usepackage{array}
\usepackage{lipsum}
\usepackage{multirow}
\usepackage{graphicx}
\usepackage{url}
\usepackage{times}

\usepackage{tikz}
\usetikzlibrary{shapes,snakes}
\usetikzlibrary{positioning,calc}
\usetikzlibrary{arrows,decorations.markings}

\usepackage{adjustbox}

\newcommand{\nop}[1]{}

\usepackage{listings}

\begin{document}
\title{Ozy: A General Orchestration Container}


\author{\IEEEauthorblockN{Glenn Osborne}
\IEEEauthorblockA{Catavolt Inc\\
Alpharetta, GA 30022, USA\\
Email: glenn.osborne@catavolt.com}
\and
\IEEEauthorblockN{Tim Weninger}
\IEEEauthorblockA{Department of Computer Science and Engineering\\
University of Notre Dame\\
Notre Dame, IN 46556\\
Email: tweninge@nd.edu}
}


%Telephone: (800) 555--1212, Fax: (888) 555--1212}


\newenvironment{blockquote}{%
  \par%
  \medskip
  \leftskip=1em\rightskip=1em%
  \noindent\ignorespaces}{%
  \par\medskip}

\maketitle

\begin{abstract}
Service-Oriented Computing is a paradigm that uses services as building blocks for building distributed applications. The primary motivation for orchestrating services in the cloud used to be distributed business processes, which drove the standardization of the Business Process Execution Language (BPEL) and its central notion that a service is a business process. In recent years, there has been a transition towards other motivations for orchestrating services in the cloud, {\em e.g.}, XaaS, RMAD. Although it is theoretically possible to make all of those services into WSDL/SOAP services, it would be too complicated and costly for industry adoption. Therefore, the central notion that a service is a business process is too restrictive. Instead, we view a service as a technology neutral, loosely coupled, location transparent procedure. With these ideas in mind, we introduce a new approach to services orchestration: Ozy, a general orchestration container. We define this new approach in terms of existing technology, and we show that the Ozy container relaxes many traditional constraints and allows for simpler, more feature-rich applications.
\end{abstract}

%TW: Abstract needs to be tighter still. What is our "missing link"?

%GO: Yes, this is the missing link: "Instead, we view a service as a technology neutral, loosely coupled, location transparent procedure."
%GO: [removed from above] Services perform procedures, which can be anything from simple information requests to complicated business processes. With core business procedures encapsulated as services, organizations are able to create new applications composed from other distributed applications.
%GO: Just to be clear, RMAD does not use an orchestration language. RMAD desperately NEEDS one. I believe it will make development much easier. The Catavolt case study is a PROBLEM study. Later this year I will begin using Ozy to improve Catavolt.
%GO: What do you think of my abstract changes, too little, good, bad?

%GO: SO HERE'S THE IDEA: The primary motivation for orchestrating services in the cloud used to be distributed business processes, hence the need for BPEL. There are now other motivations for orchestrating services in the cloud: XaaS, rich user interface, RMAD, modernizing with service micro-containers. Although it is theoretically possible to make all of those services look like WSDL/SOAP services, it would be overly pedantic and never accepted in industry. There has to be an easier way, thus Ozy. It could be viewed that with Ozy we are returning back to the original notion of SOC, programming-in-the-large over disparate modules with a "module interconnect language".

%^ok


% \fi
\IEEEpeerreviewmaketitle

\section{Introduction}

Service-Oriented Computing (SOC) is a paradigm that uses services as building blocks for building distributed applications. Services perform procedures, which can be anything from simple information requests to complicated business processes. The original focus of SOC was the automation of business processes by coordinating the services of multiple business partners. In recent years, however, the Web has evolved into an ecosystem of Everything-as-a-Service (XaaS)~\cite{duan2015everything}; consequently, services orchestration has moved beyond the need for just coordinating business processes and has evolved into a more general problem of programming-in-the-large. 

The basic Service-Oriented Architecture (SOA) is a relationship involving three kinds of participants: Service Provider, Service Client and Service Registry. The Extended Service-Oriented Architecture (ESOA) extends the basic SOA with three tiers that address overarching concerns. The middle tier is a service composition layer that facilitates orchestrations over service definitions and service implementations~\cite{papazoglou2003service}.

The standard language for service orchestration is the Business Process Execution Language (WS-BPEL or BPEL), which is based on the specifications for SOAP, WSDL and UDDI~\cite{alves2web}. Its design is centered on the notion of the business process being the glue between interacting services. 

In recent years, RESTful (REpresentational State Transfer) services have emerged as a lightweight alternative to SOAP services. Whereas SOAP services are based on remote procedure calls, RESTful services build upon the HTTP hypermedia standards of resources, resource identifiers and representations. The key elements of RESTful services are technology neutral components, connectors and resource identifiers~\cite{fielding2000architectural}.

The standard technique for composing RESTful Web services is the "mashup". There are roughly three kinds of mashups: data-oriented, process-oriented and consumer-oriented. Unlike data and consumer-oriented mashups, which permeate the Web, process-oriented mashups are less frequent, but most typically provide for the composition of business processes~\cite{garriga2016restful}.

There are several forces driving the proliferation and heterogeneity of service-oriented computing. The adoption of utility computing on a pay-as-you-go basis has allowed enterprises to develop applications more easily, cost effectively, and reliably than ever before~\cite{ranjan2015note}. New service platforms, such as Rapid Mobile App Development (RMAD)~\cite{marketguide2014RMAD}, facilitate opportunistic software development by business analysts. Finally, the need to modernize legacy systems has led to the micro-container concept that uses services to provide accessibility and increased scalability of older software architectures~\cite{yangui2015spd}.

\begin{figure}[ht]
    \centering
    \resizebox{.45\textwidth}{!}{
        \begin{tikzpicture}
\node [cloud, fill=blue!02, draw,cloud puffs=12,cloud puff arc=140, aspect=2, inner ysep=7em] {};
\node[draw, ultra thick, fill=white, inner sep=.8em, rectangle, font=\large] (catavolt) {\textbf{RMAD}};

\node[draw, thick, fill=white, rectangle, align=center, minimum height=.8cm,minimum width=2.6cm, inner sep=0cm] (sap) at (-3.1,-5.5) {SAP}; 
\node[draw, thick, fill=white, rectangle, align=center, minimum height=.8cm,minimum width=2.6cm, inner sep=0cm] (erp) at (0,-5.5) {Legacy ERP}; 
\node[draw, thick, fill=white, rectangle, align=center, minimum height=.8cm,minimum width=2.6cm, inner sep=0cm] (dataextraction) at (3.1,-5.5) {Data Extraction\\\{external use only\}}; 

\node[draw, thick, fill=white, rectangle, align=center, minimum height=.8cm,minimum width=2.6cm, inner sep=0cm] (media) at (-5.0,1) {Media}; 
\node[draw, thick, fill=white, rectangle, align=center, minimum height=.8cm,minimum width=2.6cm, inner sep=0cm] (bigdata) at (-5.0,-1) {Big Data}; 
\node[draw, thick, fill=white, rectangle, align=center, minimum height=.8cm,minimum width=2.6cm, inner sep=0cm] (social) at (5.0,1) {Social Networking\\for Business}; 
\node[draw, thick, fill=white, cylinder, shape border rotate=90, aspect=0.2, align=center, minimum height=1cm,minimum width=2.6cm, inner sep=0cm] at (5.0,-1) (metadata) {Metadata}; 

\node[draw, thick, rectangle, align=center, minimum height=.8cm,minimum width=2.6cm, inner sep=0cm] (phone) at (-3.1,5.5) {Phone}; 
\node[draw, thick, rectangle, align=center, minimum height=.8cm,minimum width=2.6cm, inner sep=0cm] (tablet) at (0,5.5) {Tablet}; 
\node[draw, thick, rectangle, align=center, minimum height=.8cm,minimum width=2.6cm, inner sep=0cm] (thing) at (3.1,5.5) {Thing}; 

\draw [thick] (catavolt) -- (media);
\draw [thick] (catavolt) -- (bigdata);
\draw [thick] (catavolt) -- (social);
\draw [thick] (catavolt) -- (metadata);
\draw [thick] (catavolt) -- (sap);
\draw [thick] (catavolt) -- (erp);
\draw [thick] (catavolt) -- (dataextraction);
\draw [thick] (catavolt) -- (phone);
\draw [thick] (catavolt) -- (tablet);
\draw [thick] (catavolt) -- (thing);

\node[draw, thick, fill=white, ellipse, minimum height=.8cm,minimum width=1.4cm] at (-2.2,.6) {File}; 
\node[draw, thick, fill=white, ellipse, align=center, minimum height=.8cm, minimum width=1.4cm, inner sep=0cm] at (-2.2,-.6) {Custom\\Driver}; 
\node[draw, thick, fill=white, ellipse, align=center, minimum height=.8cm,minimum width=1.4cm, inner sep=0cm] at (2.2,.6) {REST}; 
\node[draw, thick, fill=white, ellipse, align=center, minimum height=.8cm,minimum width=1.4cm, inner sep=0cm] at (2.2,-.6) {JDBC}; 
\node[draw, thick, fill=white, ellipse, align=center, minimum height=.8cm,minimum width=1.4cm, inner sep=0cm] at (-1.6,-2.5) {BAPI}; 
\node[draw, thick, fill=white, ellipse, align=center, minimum height=.8cm,minimum width=1.4cm, inner sep=0cm] at (0,-2.5) {SOAP}; 
\node[draw, thick, fill=white, ellipse, align=center, minimum height=.8cm,minimum width=1.4cm, inner sep=0cm] at (1.6,-2.5) {JDBC}; 
\node[draw, thick, fill=white, ellipse, align=center, minimum height=.8cm,minimum width=1.4cm, inner sep=0cm] at (0,2.5) {UI Services\\Custom HTTP/JSON}; 

\end{tikzpicture}
    }
    \caption{Heterogeneous services on an enterprise system. The Ozy container aims to elegantly orchestrate these diverse services.}% We use the system at Catavolt Inc. as an example.} %case study.}%TW - folks probably won't know what catavolt is, its important to keep this as a case study and not an advertisement.
    \label{fig:catavoltrmad}
\end{figure}
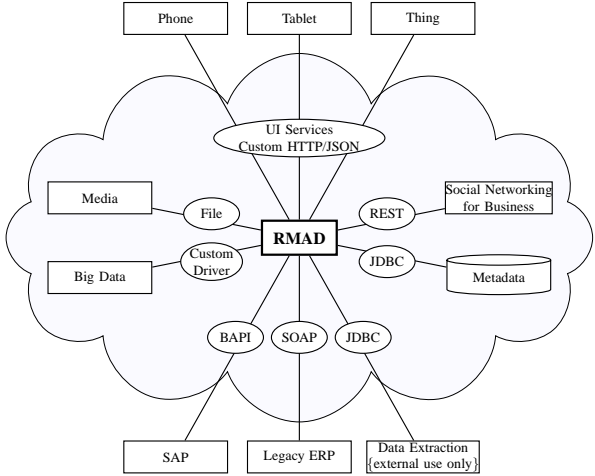

The Web service concept has evolved and now means more than just a business process. As a result, we believe that the industry can benefit from a more general treatment of services orchestration. Therefore, in this paper we introduce a new approach to services orchestration with Ozy, a general orchestration container based on the Oz Computation Model (OCM)~\cite{van2004concepts}. In addition to the description provided in the present work we have released the orchestration framework as an open source software project under the Affero GPL at \url{https://github.com/ozyio}.

To illustrate this new approach, its benefits and its restrictions, we show examples of an RMAD platform for both the development and deployment of mobile apps. As illustrated in Fig.~\ref{fig:catavoltrmad}, large enterprises typically orchestrate several kinds of services including:

\begin{itemize}
    \item Swagger/RESTful services
    \item WSDL/SOAP service
    \item Database services (NOSQL and JDBC)
    \item File services for serving media data
    \item Legacy services to access SAP (BAPI) or other older ERP services %\footnote{Although standard web services are available for SAP, a large number of customers have not upgraded and still rely on BAPI.}
    \item Proprietary HTTP/JSON services to support rich user experiences~\cite{osborne2013method}
\end{itemize}

In its original sense, a service is simply a technology neutral, loosely coupled, location transparent procedure~\cite{papazoglou2003service}. However, the prevailing techniques for orchestrating services are lacking support for the heterogeneity illustrated here. BPEL, for example, has demonstrated success in orchestrating business processes, but its central notion that a service is a business process is too restrictive. Furthermore, BPEL suffers from two strong criticisms: (1) it lacks a formal specification, and (2) its language is based on a cumbersome XML-based syntax~\cite{medeiros2014survey}. RESTful orchestration techniques are also too restrictive because they also focus on business processes~\cite{pautasso2009restful}. Moreover, the RESTful orchestration techniques that do exist are immature~\cite{garriga2016restful}.

In the present work we present Ozy, a general orchestration container. Ozy is based on the Oz Computation Model (OCM); although the Oz language is best known for teaching multi-paradigm programming practices~\cite{van2002teaching}, the underlying theory and constructs of the OCM are especially useful to services orchestration. Specifically, we exploit implicit synchronization, dataflow variables~\cite{smolka1995oz}, lazy execution, failed values and partial termination~\cite{van2004concepts}. Furthermore, the expressibility of the Oz language allows us to use programming techniques not typically found in orchestration languages, such as declarative functional programming and pattern matching~\cite{muller1995multiparadigm}. To the best of our knowledge, Ozy is the only orchestration engine that supports both implicit synchronization and persistent execution state. In addition, the Ozy orchestration container also includes the following contributions:

\begin{enumerate}
    \item A new notion of partial activation as a dual to the existing notion of partial termination
    \item A technology neutral orchestration architecture inspired by the elements of a network-based architecture
    \item An open source orchestration container that implements the orchestration architecture (2), implements an Oz language interpreter based on the OCM, and exploits the notions of partial activation (1) and partial termination to support a persistent execution state
\end{enumerate}

This paper describes the Ozy framework with special attention paid to the limitations of current approaches. After a section on related work and a brief introduction to the Oz computation model, the Ozy architecture and container systems are described. We finally provide two illustrative examples that highlights the Ozy orchestration container in real world scenarios.

\section{Related Work}

Service-Oriented Computing (SOC) is a paradigm that uses loosely-coupled services as the fundamental elements for building applications. SOC relies on the Services-Oriented Architecture (SOA) and the Extended Services-Oriented Architecture (ESOA) to define standards for publishing, finding, binding and composing services~\cite{papazoglou2003service}. Service-Oriented Middleware (SOM) supports the service-oriented interaction pattern with the provisioning of software features for deploying, publishing, finding, and binding services at runtime~\cite{qilin2010state,al2012service}.

Composing services is an exercise in programming-in-the-large. DeRemer and Kron argue that programming-in-the-large is a distinct and different intellectual activity from programming-in-the-small, and consequently, a different "module interconnect language" should be used instead of a common programming language, such as one typically used to implement modules. 

In the traditional view, there are three service composition models: Choreography, Orchestration and Wiring. Orchestration is the most used composition model, and the de facto standard is BPEL, an orchestration language based on the base specifications of SOAP, WSDL and UDDI~\cite{alves2web}. Although BPEL is the industry standard, it suffers from two big criticisms: (1) it lacks formalization and (2) is based on a cumbersome XML syntax~\cite{medeiros2014survey}. 

SOCK and JOLIE are two more-recent contributions that address some of the deficiencies of BPEL. SOCK (Service-Oriented Computing Kernel) formalizes the basic mechanisms of services communications and composition, therefore addressing the lack of formalism in BPEL~\cite{guidi2006sock}. JOLIE (Java Orchestration Language Interpreter Engine) builds on SOCK to create an expressive language and orchestration interpreter engine, therefore addressing the cumbersome XML-based syntax of BPEL~\cite{montesi2007jolie,montesi2007composing}.

The open source Apache ODE (Orchestration Director Engine) project is the underlying execution engine for several BPEL implementations, such as JBoss RiftSaw and WSO2 Business Process Server. Apache ODE relies on the Apache JaCOb (Java Concurrent Objects) framework to implement two key features required of BPEL processes: concurrency and persistence of execution state.

In recent years, and because of the success of the RESTful architecture, a Web service is more generally defined as a SOAP-based or RESTful-based service that can be described, advertised and discovered through Internet-based protocols~\cite{sheng2014web}. SOAP services are based on remote procedure calls. In contrast, RESTful services build on the HTTP hypermedia standards of resources, resource identifiers and representations. The key elements of RESTful services are technology neutral components, connectors and resource identifiers~\cite{fielding2000architectural}.

Lightweight RESTful services are designed to ease consumption, composition and have been successfully used to build community-driven services known as mashups~\cite{de2011building,garriga2016restful}. Another approach for composing RESTful services is a technique that uses BPEL to orchestrate RESTful services by wrapping RESTful services with WSDL definitions~\cite{pautasso2009restful}.

Cloud computing has emerged in conjunction with Web services as a dominant utility computing solution. Cloud vendors offer platform-as-a-service (PaaS) and software-as-a-service (SaaS) applications on a pay-as-you-go basis allowing enterprises to develop applications more easily, cost effectively, and reliably than ever before~\cite{ranjan2015note}. Cloud services have become so prolific that we now refer to the provider ecosystem as Everything-as-a-Service (XaaS)~\cite{duan2015everything}.

Opportunistic software methods are being used in cloud computing to drive down the backlog of applications by allowing non-traditional software developers to combine two or more software systems to create functionality that mainstream software does not provide~\cite{balasubramaniam2008situated}. For example, Gartner recently identified the Rapid Mobile App Development (RMAD) category of development tools~\cite{marketguide2014RMAD}. RMAD is opportunistic software development for mobile apps that offers simpler, faster development by business analysts. %An industry example of an opportunistic RMAD system is used by Catavolt Inc., a PaaS/SaaS platform for both the development and deployment of mobile apps.

The need to reduce application backlogs is just one of many motivating forces affecting service-oriented computing in the cloud. Service-oriented computing has focused primarily on easy integration at the server-side at the price of lower flexibility and ease-of-use at the client-side~\cite{atkinson2012unified}. Focus is shifting to client-side interactions with services that sometimes require innovative service protocols~\cite{osborne2013method}. Another force affecting cloud computing is the need to modernize legacy software. The micro-container concept uses services to provide accessibility and increased scalability of older software architectures~\cite{yangui2015spd}.
%GO: Micro-services (a.k.a services micro-container) is a very hot topic right now for cloud computing used to scale line-of-business software in the cloud by creating lots of servers. The trick is achieve scalability by creating more server instances, redundantly, for the more active services.

%GO: SO HERE'S THE IDEA: The primary motivation for orchestrating services in the cloud used to be distributed business processes, hence the need for BPEL. There are now other motivations for orchestrating services in the cloud: XaaS, rich user interface, modernizing with service micro-container.
\subsection*{The Missing Link}
%TW - punch them in the face with it.

\begin{table}[t]
    \caption{Feature comparison of existing orchestration languages: BPEL, JOLIE and the present work Ozy. Oz, itself, is not an orchestration language, but is included here for comparison.}
    \label{tab:comparison}
\centering
    \begin{tabular}{ l | c | c | c | c }
        & BPEL & JOLIE & Oz & \textbf{Ozy} \\ \hline
        Formal Specification & - & + & + & + \\ \hline
        High-Level Language & - & + & + & + \\ \hline
        Interpreted & + & + & - & + \\ \hline
        Functional Programming & - & - & + & + \\ \hline
        Long Running & + & + & - & + \\ \hline
        Declarative Concurrency & + & + & + & + \\ \hline
        Implicit Synchronization & - & - & + & + \\ \hline
        Pattern Matching & - & - & + & + \\ \hline
        Unification & - & - & + & + \\ \hline
        Entailment & - & - & + & + \\ \hline
    \end{tabular}
    \vspace{-.5cm}
\end{table}

Composing services, and orchestrating in particular, has moved beyond the need for coordinating business processes and has become a more general programming-in-the-large problem. Although the Oz programming language is best known for its support of multi-paradigm programming~\cite{van2002teaching}, the theory and constructs of the Oz Computation Model (OCM) are especially useful for services orchestration. Most notable are implicit synchronization, dataflow variables~\cite{smolka1995oz}, lazy execution, failed values and partial termination~\cite{van2004concepts}. Furthermore, the expressibility of the Oz language enables programming techniques not typically found in programming-in-the-large~\cite{muller1995multiparadigm}. In the present work, we adapt the Oz Computational Model for practical use in service orchestration. We call this adaption Ozy, and show that it is a more elegant, feature rich orchestration container.

%GO: Can you use the following to fix this and other areas of the document where you need to differentiate between Oz and Ozy?

%GO: Ahh, okay. See the "Languages by Feature" table, it probably belongs earlier. Furthermore, Ozy is an embeddable orchestration engine. Specifically, it's an Oz language interpreter meant to be embedded in other high-level languages to reuse the underlying procedure libraries. Also, the Ozy runtime is founded on non-blocking asynchronous message-passing, Oz is not. It's a little confusing because you can implement non-blocking asynchronous message-passing in Oz, but it is not FOUNDED on it. Our first implementation is in Java. We have plans to implement it in C# for the Microsoft CLR. The differences gained with Ozy: is embeddable in other languages, interoperable with host libraries, founded on message routing, uses interpretive computation, and can persist execution state. Is that enough? :)

%TW I can make due with this.
 
\section{The Oz Computation Model}

The Oz Computation Model (OCM) \cite{van2004concepts} contains abstractions especially useful for services orchestration. Specifically, we exploit \textit{implicit synchronization}, \textit{dataflow variables}~\cite{smolka1995oz}, \textit{lazy execution}, \textit{failed values} and \textit{partial termination}~\cite{van2004concepts} provided by the OCM. Furthermore, the expressibility of the Oz language allows advanced programming techniques, such as declarative functional programming and pattern matching~\cite{muller1995multiparadigm}.

The \textit{dataflow variable} is the most powerful OCM abstraction because it greatly simplifies concurrency by hiding explicit synchronization, thereby giving us \textit{implicit synchronization}. Dataflow variables provide implicit synchronization at the most atomic level--a single variable. A dataflow variable is created and bound in separate steps. The single rule of accessing a dataflow variable is very simple: \textit{execution waits until the dataflow variable is bound. Once bound, execution continues}. This property allows advanced programming techniques never before seen in an orchestration language. The power of dataflow variables is best explained by Smolka: 

\begin{blockquote}
A basic problem with existing programming languages is that they delegate the creation and coordination of concurrent computational activities to the underlying operating system and network protocols. This has the severe disadvantage that the \textit{data abstractions of the programming language cannot be shared between communicating computational agents}. Thus the benefits of existing programming languages do not extend to the central concerns of concurrent and distributed software systems~\cite{smolka1995oz} (emphasis added).
\end{blockquote}

The present work, Ozy, uses the dataflow variable abstraction thereby allowing service messages to be shared among interacting services independent of the underlying operating system and network protocols. 

For the following examples on dataflow variables, the complete Oz syntax can be found in \cite{van2004concepts}. Our simple example, illustrated in Fig.~\ref{fig:ozexample1}, executes 3 concurrent threads. The third thread waits on variables $X$ and $Y$ before continuing. This is an example of implicit synchronization because the third thread is void of any syntax that explicitly waits for variables $X$ and $Y$ to be bound.

\begin{figure}[ht]
    %\centering
\hspace*{.25cm}
\begin{tikzpicture}
  \node(listing1) {
\begin{lstlisting}
thread X = 5 end
thread Y = 7 end
thread Z = X + Y end
\end{lstlisting}
};
%\node[draw, ultra thick, fill=white, inner sep=.8em, rectangle, font=\large] (catavolt) 
\end{tikzpicture}

    \caption{Simple example of implicit synchronization with 3 threads. Despite the lack of explicit syntax, implicit synchronization via dataflow variables requires that the third thread waits for variables $X$ and $Y$ to be bound ({\em i.e.}, thread 1 and 2 to complete) before continuing.}
    \label{fig:ozexample1}
\end{figure}

A more sophisticated example of dataflow programming illustrated in Fig.~\ref{fig:ozexample2} involves nested data structures and function calls\footnote{In Oz, function signatures and function applications are delimited with curly braces.}. Dataflow variables can be arbitrarily nested in both data structures and logic. Because synchronization is implicit, logic can be written very cleanly, void of explicit techniques that synchronize concurrency.

\begin{figure}[ht]
    %\centering
    \hspace*{.25cm}
\begin{tikzpicture}
  \node(listing1) {
\begin{lstlisting}
fun {GetCustomerInfo Id}
  fun {GetAverageSale Id} Answer in
    thread
      {Sleep 3 days}
      Answer = 12345.67
    end
    return Answer
  end
  fun {GetLastSale Id}
    123
  end
  customer(key:Id averageSale:{GetAverageSale Id} lastSale:{GetLastSale Id})
end
local C D in 
  C = {GetCustomerInfo 'acme'}
  case C of customer(key:Id averageSale:AS lastSale:LS)
    D = if LS > AS then 'Y' else 'N' end
  end
end
\end{lstlisting}
};

\node[draw, text=white, thick, fill=blue!60, circle, font=\scriptsize, inner sep=.75mm] (1) at (-3.5,-2.7) {1} ;
\node (ls) at (-4.0,-2.1) {};
\node (as) at (-3.0,-2.1) {};

\draw [->, thick] (1) to (as);
\draw [->, thick] (1) to (ls);

\node[draw, text=white, thick, fill=blue!60, circle, font=\scriptsize, inner sep=.75mm] (2) at (1.3,-1.3) {2} ;
\node (as) at (0.4,-1.7) {};
\node (ls) at (2.2,-1.7) {};

\draw [->, thick] (2) to (as);
\draw [->, thick] (2) to (ls);

\node[draw, text=white, thick, fill=blue!60, circle, font=\scriptsize, inner sep=.75mm] (3) at (-1.8,0.2) {3} ;
\node (customer) at (-4.4,-0.28) {};
\draw [->, thick] (3) to (customer);

\draw [decorate,decoration={brace,amplitude=5pt},xshift=-4pt,yshift=0pt]
(-0.9,-0.2) -- (1.6,-0.2) node [draw, text=white, thick, fill=blue!60, circle, font=\scriptsize, inner sep=.75mm ,midway, yshift=12pt] {4};

\node[draw, text=white, thick, fill=blue!60, circle, font=\scriptsize, inner sep=.75mm] (5) at (-1.4,1.0) {5} ;
\node (answer) at (-3.15,1.0) {};
\draw [->, thick] (5) to (answer);

\node[draw, text=white, thick, fill=red, circle, font=\scriptsize, inner sep=.75mm] (6) at (-1,2) {6} ;
\node (thread) at (-4.2,2.1) {};
\node (days) at (-2.7,1.85) {};

\draw [->, thick] (6) to (thread);
\draw [->, thick] (6) to (days);

\end{tikzpicture}
    \caption{Sophisticated example of dataflow programming that computes a discount $D$ based on a complex, concurrent and implicitly synchronized work flow. Existing orchestration languages are ill-equipped to elegantly orchestrate this process.}
    \label{fig:ozexample2}
    \vspace{-.5cm}
\end{figure}

%\begin{figure}[ht]
  %  \centering

 %   \includegraphics[width=.95\linewidth]{OzExample2}
 %   \caption{Example 2}
 %   \label{fig:ozexample2b}
%\end{figure}

Although our sophisticated example appears intuitive, there are actually deep dependencies on concurrency and synchronization: (1) two numeric values are compared that (2) depend on the result of pattern matching that (3) depends on the result of a function call to \texttt{GetCustomerInfo} that constructs a compound value that depends on two other nested function calls to \texttt{GetAverageSale} and \texttt{GetLastSale} where (4) one of the nested function calls depends on (5) an answer that (6) is fulfilled concurrently and takes three days to complete. Dataflow variables greatly simplify concurrent programming. Without dataflow variables, this example written in a mainstream programming language would require explicit synchronization or a concurrency library, such as monadic combinators for a functional language or an actor library for an imperative language. %TW--<- Maybe cut the final phrase (such as...) to save space later.

Another key OCM abstraction is the Kernel Language, a concise set of values and statements that serve as the basic computational elements and as building blocks for higher level computing techniques. The OCM defines formal operational semantics and an abstract machine that specify exactly how to execute a kernel language program. Informally, the abstract machine is a process consisting of several semantic stacks and a shared store of dataflow variables. Each semantic stack corresponds to a thread, and all semantic stacks access the same store. Execution begins with an initial process state and proceeds as a sequence of state transitions where a semantic stack (thread) is chosen and reduced. Processing continues until all semantic stacks are null. Semantic stacks can be reduced in parallel with synchronization on the store of dataflow variables. 

More formally, the Oz computation model is based on the following definitions:
\begin{itemize}
    \item[] $\sigma$ = single-assignment store (dataflow variables)
    \item[] $s$ = kernel statement
    \item[] $E$ = environment
    \item[] $(s, E)$ = semantic statement
    \item[] $[(s, E)]$ = semantic stack (ST)
    \item[] $\{[(s, E)]\}$ = multiset semantic stack (MST)
\end{itemize}
A computation step begins by choosing from the multiset one runnable (reducable) semantic stack giving us the following definition:
\begin{itemize}
    \item[(a)] MST = \{ST\} $\uplus$ MST$^\prime$
\end{itemize}
With a semantic stack chosen, a computation step is completed with the following definition:
\begin{itemize}
    \item[(b)] $(\{\textrm{ST}\} \uplus \textrm{MST}^\prime, \sigma) \rightarrow (\{\textrm{ST}^\prime\} \uplus \textrm{MST}^\prime, \sigma^\prime)$
\end{itemize}
%TW what does the prime mean?
Program execution repeats steps (a) and (b) until termination~\cite{van2004concepts}.
 
The OCM executes over dataflow variables, which leads to the notion of \textit{partial termination}, a concept useful for reasoning about process state. To explain partial termination, let us define a \textit{list} as a pair (\textit{head}, \textit{tail}), where \textit{head} is any value and \textit{tail} is a \textit{list} value or \textit{null}. An input stream is simply a list with an unbound tail. If processing an input stream with an unbound tail, the program may never terminate. However, if the input stream stops growing, then based on the single rule for a dataflow variable (see above), the program will eventually stop executing while it waits for the input stream to grow ({\em i.e.}, for a new tail to be bound). This is an example of partial termination. It has not terminated completely because if the input stream grows, the program will execute further, up to the next partial termination. If there are no further input, then the program will execute no further. Partial termination is particularly useful because it allows us to determine that a long running process is waiting on an external message.

\nop{%TW nop is the block comment command, this means that I'm trying to cut.
\begin{figure}[ht]
    \centering
    \resizebox{.22\textwidth}{!}{
        \begin{tikzpicture}

\node [draw, ultra thick, circle, fill=blue!05!white, minimum size=60, align=center] (active) at (0,0) {Partially\\Active};
\node [draw, ultra thick, dashed, circle, fill=blue!05!white, minimum size=60, align=center] (term) at (0,-4) {Partially\\Terminated};

\node [draw, circle, fill=black, minimum size=20] (init) at (0,3.5) {};
\node [draw, circle, ultra thick, fill=white, minimum size=35] (endouter) at (3.5,0) {};
\node [draw, circle, fill=black, minimum size=20] (end) at (3.5,0) {};

\draw[very thick, decoration={markings,mark=at position 1 with
    {\arrow[scale=2,>=latex]{>}}},postaction={decorate}] (init) -- (active) node [midway, left, fill=white] {1};
    
\draw[very thick, decoration={markings,mark=at position 1 with
    {\arrow[scale=2,>=latex]{>}}},postaction={decorate}] (active) -- (term) node [midway, left, fill=white] {2};
   
\draw[very thick, decoration={markings,mark=at position 1 with
    {\arrow[scale=2,>=latex]{>}}},postaction={decorate}]  (term.west) -- +(-1,0) --  node [midway, left, fill=white] {3} +(-1,4) -- (active.west) ;

\draw[very thick, decoration={markings,mark=at position 1 with
    {\arrow[scale=2,>=latex]{>}}},postaction={decorate}] (active) -- (endouter) node [midway, above, fill=white] {4};
    
\end{tikzpicture}
    }
    \caption{Long-running process life-cycle with partially actice and partially terminated states.}
    \label{fig:processlifecycle}
\end{figure}
}

The new notion of \textit{partial activation} occurs when a new semantic stack is added to a partially terminated process, therefore guaranteeing at least one semantic reduction. The process is then partially activated, which means that it will execute, but it may or may not execute to complete termination as described above. Generalizing partial activation allows for a variety of activation scenarios in addition to the scenario above that uses a stream. Partial activation combined with partial termination allows us to define the life-cycle model for long-running processes\nop{ shown in Fig.~\ref{fig:processlifecycle}}. The life-cycle for long-running processes behaves as follows:
\begin{enumerate}
    \item The process is created by initializing an abstract machine with an initial semantic stack that is reducible. The process is now partially activated and execution begins.
    \item The process waits on an external message and cannot execute further because all semantic stacks are waiting on dataflow variables and therefore cannot be reduced. The process is now partially terminated and eligible to be saved to disk.
    \item An external message arrives for a partially terminated process. The process is restored, from disk if necessary, and a new semantic stack is added to the process. The process is now partially activated and execution begins.
    \item The process has been fully reduced, all semantic stacks are empty. The process is now completely terminated.
\end{enumerate}

\textit{Lazy execution} and \textit{failed values} are two other OCM constructs important to services orchestration. Lazy execution allows for late binding of modules and supports advanced functional programming techniques. Late binding is crucial to large process configurations so that services are bound only when necessary. Failed values are necessary to properly communicate failures. When the computation of a dataflow variable fails, a failed value is stored in the dataflow variable and the failure is propagated to other threads waiting on that dataflow variable.

In summary, so far we have reviewed the Oz Kernel Language and Oz Abstract Machine, as well as the highlights of the theory behind the Oz Computational Model that are most relevant to services orchestration: implicit synchronization, dataflow variables, lazy execution and failed values. These constructs combined with the new notion of partial activation form the technical foundation for the Oz language interpreter presented later in this paper. %TW maybe we need to be more explicit about Ozy vs Oz here and earlier in the paper.

\section{Ozy Architecture}
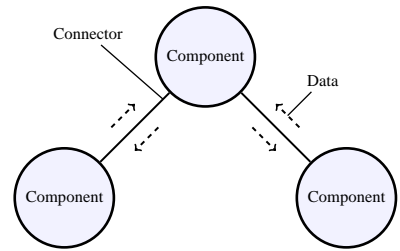
\begin{figure}[ht]
    \centering
    \resizebox{.30\textwidth}{!}{
        \begin{tikzpicture}

\node [draw, ultra thick, circle, fill=blue!05!white, minimum size=60, align=center] (top) at (0,0) {Component};
\node [draw, ultra thick, circle, fill=blue!05!white, minimum size=60, align=center] (left) at (-3,-3) {Component};
\node [draw, ultra thick, circle, fill=blue!05!white, minimum size=60, align=center] (right) at (3,-3) {Component};

\draw [very thick] (top) -- (left);
\draw [very thick] (top) -- (right);

\draw [very thick, ->, dashed] (2.0,-1.5) -- (1.5,-1.0);
\draw [very thick, ->, dashed] (1.0,-1.5) -- (1.5,-2.0);

\draw [very thick, ->, dashed] (-2.0,-1.5) -- (-1.5,-1.0);
\draw [very thick, ->, dashed] (-1.0,-1.5) -- (-1.5,-2.0);

\node [align=center] (conn) at (-2.5,0.5) {Connector};
\node [align=center] (data) at (2.5,-0.5) {Data};

\draw [thick] (data) -- (1.75,-1.25);
\draw [thick] (conn) -- (-0.9,-0.9);

\end{tikzpicture}
    }
    \caption{Elements of the Network-Based Architecture (net-arch) wherein components interact by sending data through communications connectors.}
    \label{fig:networkarchelements}
\end{figure}

%TW We definitely need a segue here. Ozy vs Oz.
The Ozy architecture is based on the general view of the network-based architecture (net-arch)~\cite{fielding2000architectural}. The conceptual elements of the net-arch are \textit{Components}, \textit{Connectors} and \textit{Data}. As illustrated in Fig.~\ref{fig:networkarchelements}, components interact by sending data through communications connectors (message passing). Components, and only components, perform computation on state, which can be a combination of data passed on connectors and private data known only to the component. The connector is especially useful because it encapsulates the technology used to communicate with other services. Components, connectors and data messages are the basic abstractions used in the Ozy architecture.

The manner in which Ozy employs the connector abstraction is what differentiates it from SOA-based and RESTful-based architectures. The SOA architecture and its BPEL are directly dependent and therefore constrained by the underlying specifications for WSDL and UDDI. The RESTful architecture is directly dependent on the uniform HTTP operations of GET, PUT, POST and DELETE. In contrast, the Ozy architecture loosens the constraints on the connector abstraction. A connector simply transfers data messages from one interface to another without any semantic transformation. Internally, the connector may be a subsystem that transforms and adapts, but when the message is delivered, it is semantically unchanged. 

We must emphasize that the Ozy architecture is designed for hiding services implementations. Ozy does not define a new kind of services interface. It is very important that public interfaces for composite services be founded on industry standards, which is currently either SOAP or RESTful style interfaces. By moving the dependencies on network interfaces into connector abstractions (adapters), Ozy allows services containers to operate freely and evolve independently as long as service invocations can be adapted between private and public interfaces. Furthermore, private-only compositions can be optimized to simply use local interfaces.    

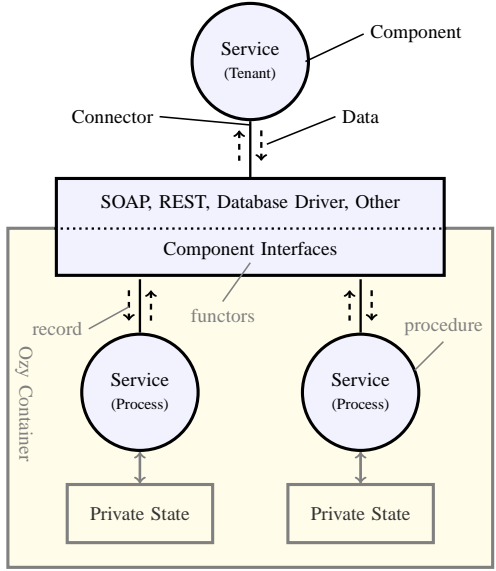
\begin{figure}[t]
    \centering
        \resizebox{.38\textwidth}{!}{
        \begin{tikzpicture}

\node [draw, ultra thick, circle, fill=blue!05, minimum size=60, align=center] (service) at (0,5) {Service \\\footnotesize{(Tenant)}};

\node [draw=gray, ultra thick, rectangle, fill=yellow!10, minimum width=250, minimum height=175, align=center] (bigbox) at (0,-1.1) {};

\node [draw, ultra thick, rectangle, fill=blue!05, minimum width=200, minimum height=50, align=center] (soap) at (0,2) {SOAP, REST, Database Driver, Other\\\\Component Interfaces};

\node [draw, ultra thick, circle, fill=blue!05, minimum size=60, align=center] (serviceleft) at (-2,-1) {Service\\\footnotesize{(Process)}};
\node [draw, ultra thick, circle, fill=blue!05, minimum size=60, align=center] (serviceright) at (2,-1) {Service\\\footnotesize{(Process)}};

\node [draw=gray, ultra thick, rectangle, minimum width=75, minimum height=30, align=center, text=black!80] (soap) at (2,-3.2) {Private State};
\node [draw=gray, ultra thick, rectangle, minimum width=75, minimum height=30, align=center, text=black!80] (soap) at (-2,-3.2) {Private State};

\draw [very thick] (service) -- (0,2.88) ;
\draw [dashed, very thick, ->] (0.2,3.75) -- (0.2,3.2);
\draw [dashed, very thick, <-] (-0.2,3.75) -- (-0.2,3.2);

\draw [gray, very thick, ->] (serviceleft) -- (-2,-2.65);
\draw [gray, very thick, ->] (-2,-2.65) -- (serviceleft);

\draw [gray, very thick, ->] (serviceright) -- (2,-2.65) ;
\draw [gray, very thick, ->] (2,-2.65) -- (serviceright) ;

\draw [very thick] (serviceright) -- (2,1.05) ;
\draw [very thick] (-2,1.05) -- (serviceleft) ;

\draw [dashed, very thick, ->] (-2.2,0.85) -- (-2.2,0.3);
\draw [dashed, very thick, <-] (-1.8,0.85) -- (-1.8,0.3);
\draw [dashed, very thick, ->] (2.2,0.85) -- (2.2,0.3) ;
\draw [dashed, very thick, <-] (1.8,0.85) -- (1.8,0.3);

\draw [dotted, very thick] (-3.55,1.97) -- (3.55,1.97);

\node [] (component) at (3,5.5) {Component};
\draw [thick] (component.west) -- (1,5);

\node [] (data) at (2,4) {Data};
\draw [thick] (data.west) -- (0.3,3.5);

\node [] (conn) at (-2.5,4.0) {Connector};
\draw [thick] (conn.east) -- (-0.01,3.85);

\node [text=gray] (record) at (-3.5,0.15) {record};
\draw [gray, thick] (record.east) -- (-2.2,0.55);

\node [text=gray] (functors) at (-0.5,0.4) {functors};
\draw [gray, thick] (functors.north) -- (0,1.35);

\node [text=gray] (proc) at (3.5,0.2) {procedure};
\draw [gray, thick] (proc.south) -- (2.98,-0.6);

\node [text=gray, rotate=-90] (cont) at (-4.1,-1.2) {Ozy Container};

\end{tikzpicture}
    }
    \caption{The Ozy Architecture. Ozy does not define a new services interface; industry standard SOAP or RESTful interfaces should still be employed, but the Ozy Architecture does provide a better separation of concern between interface and implementation.}
    \label{fig:ozyorcharch}
    \vspace{-.5cm}
\end{figure}

Ozy exploits the conceptual elements of the net-arch to create a better separation of concern between service interfaces and service implementations. As shown in Fig.~\ref{fig:ozyorcharch}, an Ozy interface can be SOAP or RESTful as long as the supporting connectors are available to adapt SOAP or RESTful messages to Oz language elements, such as functors, records and procedures. Likewise, the Ozy implementation can invoke any service; public or private, local or remote, given that the supporting connectors can translate Oz language elements into target interfaces. Finally, non-standard connectors can be installed to transparently support interactions with non-standard protocols, such as NoSQL drivers and legacy ERP interfaces.

\section{Ozy Container}

The Ozy container is implemented in accordance with the Ozy architecture. The container hosts two fundamental activities: message routing and computation. The connector abstraction is realized by the computations that are performed along the message paths. When a message reaches a process interface, Ozy invokes an Oz language interpreter to compute the message. 

Ozy employs an actor model for non-blocking asynchronous message routing. There are two types of messages: \textit{Tell} and \textit{Ask}. Unlike Tell, the Ask message requires a response and is built on an asynchronous Future/Promise library that routes responses back to the requester. 

\begin{figure}[t]
    \centering
        \resizebox{.38\textwidth}{!}{
        \begin{tikzpicture}

%\node [draw=gray, ultra thick, rectangle, fill=yellow!10, minimum width=200, minimum height=220, align=center] (bigbox) at (0,-1.1) {};
%\node [draw=gray, ultra thick, rectangle, fill=blue!05, minimum width=155, minimum height=185, align=center] (bigbox) at (-0.4,-1.1) {};
%\node [draw=gray, ultra thick, rectangle, fill=blue!05, minimum width=20, minimum height=185, align=center] (bigbox) at (2.8,-1.1) {};

%\node [] (ozy) at (-3,2.4) {Ozy};
%\node [] (routing) at (-2.3,1.8) {Routing};

\node [] (in) at (-1.6,2.2)  (in) {Message In};
\node [] (out) at (2.0,2.2) (out) {Message Out};

\draw [dashed, ultra thick, ->] (-0.2,2.0) -- (-0.2,1.35);
\draw [dashed, ultra thick, <-] (0.4,2.0) -- (0.4,1.35);

\draw [] (in) -- (-0.4,1.75);
\draw [] (out) -- (0.6,1.75);

\node [draw, circle, fill=blue!09, inner sep=2] (r) at (0.1,1.0) {R};
\draw [thick] (r.north) -- (0.1,1.9);
\draw [thick, dotted] (0.1,1.9) -- (0.1,2.3);

\node [draw, circle, fill=blue!09, inner sep=2] (tl) at (-1.3,0.0) {T};
\node [draw, circle, fill=blue!09, inner sep=2] (tc) at (0.1,0.15) {T};
\draw [gray, dotted, very thick] (tc) -- (0.1, -0.3);
\node [draw, circle, fill=blue!09, inner sep=2] (tr) at (1.5,0) {T};

\draw [gray, very thick] (r) -- (tl);
\draw [gray, very thick] (r) -- (tc);
\draw [gray, very thick] (r) -- (tr);

\node [draw, circle, fill=blue!09, inner sep=2] (pl) at (-2.3,-1.0) {P};
\draw [gray, dotted, very thick] (pl) -- (-2.3, -1.5);
\node [draw, circle, fill=blue!09, inner sep=2, dashed] (plc) at (-2.0,-1.8) {P};
\node [draw, circle, fill=blue!09, inner sep=2] (pc) at (-1.3,-1.0) {P};
\draw [gray, dotted, very thick] (pc) -- (-1.3, -1.5);
\node [draw, circle, fill=blue!09, inner sep=2] (prc) at (-0.5,-1.8) {P};
\node [draw, circle, fill=blue!09, inner sep=2] (pr) at (-0.3,-1.0) {P};
\draw [gray, dotted, very thick] (pr) -- (-0.3, -1.5);

\draw [gray, very thick] (tl) -- (pl);
\draw [gray, very thick] (tl) -- (plc);
\draw [gray, very thick] (tl) -- (pc);
\draw [gray, very thick] (tl) -- (prc);
\draw [gray, very thick] (tl) -- (pr);

\node [draw, circle, fill=blue!09, inner sep=2] (prr) at (1.5,-1.0) {P};
\draw [gray, dotted, very thick] (prr) -- (1.5, -1.5);

\draw [gray, very thick] (tr) -- (prr);

\node [draw, circle, fill=blue!09, inner sep=1, dashed] (thll) at (-2.0,-3.5) {Th};
\node [draw, circle, fill=blue!09, inner sep=1, dashed] (thlc) at (-2.4,-2.8) {Th};
\node [draw, circle, fill=blue!09, inner sep=1, dashed] (thlr) at (-1.6,-2.8) {Th};

\draw [gray, very thick] (plc) -- (thll);
\draw [gray, very thick] (plc) -- (thlc);
\draw [gray, very thick] (plc) -- (thlr);

\node [draw, circle, fill=blue!09, inner sep=1] (thrl) at (-0.9,-2.8) {Th};
\node [draw, circle, fill=blue!09, inner sep=1] (thrr) at (-0.1,-2.8) {Th};

\draw [gray, very thick] (prc) -- (thrr);
\draw [gray, very thick] (prc) -- (thrl);

%\node [text=black, rotate=-90] (cont) at (2.8,-1.1) {Runtime Control};

\node [draw, circle, fill=blue!09, inner sep=1.5] (r) at (-3.9,-4.5) {R};
\node [] (r) at (-3.2,-4.5) {Root};
\node [draw, circle, fill=blue!09, inner sep=1.5] (r) at (-2.1,-4.5) {T};
\node [] (r) at (-1.1,-4.5) {Tenant};
\node [draw, circle, fill=blue!09, inner sep=1.5] (r) at (0.3,-4.5) {P};
\node [] (r) at (1.2,-4.5) {Process};
\node [draw, circle, fill=blue!09, inner sep=0.1] (r) at (2.5,-4.5) {Th};
\node [] (r) at (3.4,-4.5) {Thread};

\node [draw, circle, fill=blue!09, minimum width=14, dashed, thick] (r) at (-3.5,-5.3) {};
\node [] (r) at (-1.6,-5.3) {Partially Terminated};

\node [draw, circle, fill=blue!09, minimum width=14, thick] (r) at (0.5,-5.3) {};
\node [] (r) at (2.3,-5.3) {Partially Activated};

\draw [gray, very thick] (-2.7, -6.1) -- (-1.9, -6.1);
\node [] (r) at (-0.9,-6.1) {Connector};

\draw [ultra thick, ->, dashed] (0.5, -6.1) -- (1.3, -6.1);
\node [] (r) at (2.2,-6.1) {Message};

\end{tikzpicture}
    }
    \caption{Message routing in Ozy. Message routing is based on the Ozy Architecture where the net-arch components are Root, Tenant, Process and Thread.}
    \label{fig:msgrouting}
    \vspace{-.5cm}
\end{figure}
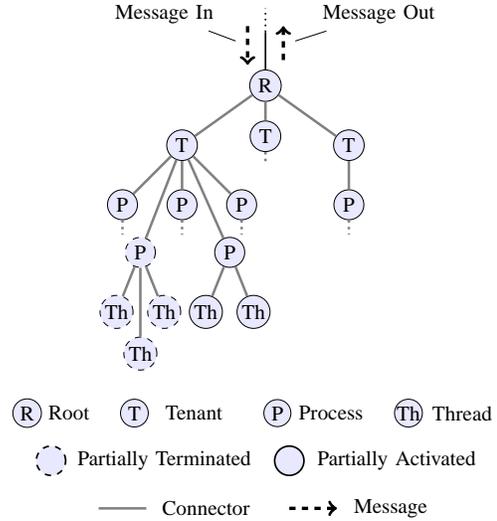

Message routing is intrinsically multi-tenant. Figure~\ref{fig:msgrouting} illustrates the message routing in Ozy, where the tenant and process services from Fig.~\ref{fig:ozyorcharch} corresponds to the tenants and processes in Fig~\ref{fig:msgrouting}. Message routing is based on the following hierarchy template: {\small \texttt{/root/tenants/\{tenantId\}/processes/\{processId\}}} where curly braces denote identifying parameters. Messages are first received in the root actor mailbox where they are subsequently processed and routed to the appropriate tenant actor. The tenant actor receives request messages in its mailbox where they are subsequently processed and routed to process actors. If necessary, a tenant actor will activate a process actor prior to sending it a message if that process has been partially terminated and had its execution state persisted to disk.

\nop{
\begin{figure}[ht]
    %\centering
\hspace*{.25cm}
\begin{tikzpicture}
  \node(listing1) {
\begin{lstlisting}
proc next_tenant_message(m) 
    if existing_process(m)
        pid = correlate_with_process_id(m)
        p = resolve_process(pid)
    else 
        p = create_process()
    end    
    next_process_message(p, m)
end
\end{lstlisting}
};
%\node[draw, ultra thick, fill=white, inner sep=.8em, rectangle, font=\large] (catavolt) 
\end{tikzpicture}
    \caption{Next Tenant Message. A message received by a Tenant must be correlated to a Process. A Process may have to be activated before processing can continue.}
    \label{fig:nexttenantmessage}
\end{figure}
}

Messages received from external services must be correlated to process instances and procedure invocations. \nop{Figure~\ref{fig:nexttenantmessage}  shows how an incoming tenant message must be correlated to an existing process.} A persistent correlation table is used to map business attributes to process identifiers. For example, suppose that a request to create a supply-order is being processed and a correlation entry is made that associates the supply-order number with the running process id. Now let us suppose the process partially terminates because the supply-order must be approved before it can be fulfilled. Later, when the supply manager approves (or disapproves) the supply-order we can use the correlation table to find the partially terminated process, activate it and allow it to continue executing. 

\begin{figure}[ht]
    %\centering
\hspace*{.25cm}
\begin{tikzpicture}
  \node(listing1) {
\begin{lstlisting}[mathescape=true]
proc next_process_message(p, m)
    (s, E) = correlate_with_semantic_stack(p, m)
    p.MST = [(s, E)] $\uplus$ p.MST 
    p.run()
end    
\end{lstlisting}
};
%\node[draw, ultra thick, fill=white, inner sep=.8em, rectangle, font=\large] (catavolt) 
\end{tikzpicture}
    \caption{Next Process Message. A message received by a Process must be correlated to a semantic statement. A typical correlation simply invokes a procedure.}
    \label{fig:nextprocessmessage}\
    \vspace{-.2cm}
\end{figure}

Similarly, correlation entries must be made to map external attribute names to dataflow variables. Figure~\ref{fig:nextprocessmessage} shows how an incoming process message must be correlated to a semantic stack. Suppose that approving a supply-order involved setting an approved-flag to true or false. Now suppose that the \texttt{createSupplyOrder} process is partially terminated and waiting on the approved-flag dataflow variable known internally, and that a correlation entry exists with this fact. Later, when the supply manager approves (or disapproves) the supply-order, the approved-flag correlation entry can be used to construct a semantic stack that will set the approved-flag dataflow variable and allow the process to continue executing.

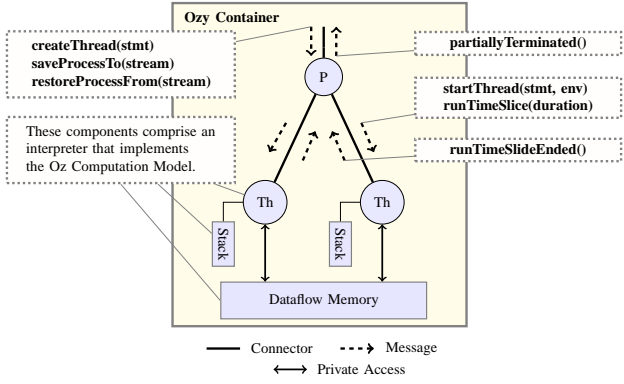
\begin{figure}[t]
    \centering
        \resizebox{.98\linewidth}{!}{
        \begin{tikzpicture}

\node [draw=gray, ultra thick, rectangle, fill=yellow!10, minimum width=200, minimum height=220, align=center] (bigbox) at (0,-1.1) {};
\node [draw=gray, ultra thick, rectangle, dotted, fill=yellow!01, minimum width=155, minimum height=45, align=left] (createthread) at (-4.75,1.3) {\textbf{createThread(stmt)}\\\textbf{saveProcessTo(stream)}\\\textbf{restoreProcessFrom(stream)}};
\node [draw=gray, ultra thick, rectangle, dotted, fill=yellow!01, minimum width=155, minimum height=45, align=left] (these) at (-4.75,-0.75) {These components comprise an\\interpreter that implements\\the Oz Computation Model.};
\node [draw=gray, ultra thick, rectangle, dotted, fill=yellow!01, minimum width=140, minimum height=15, align=left] (pterm) at (4.75,1.8) {\textbf{partiallyTerminated()}};
\node [draw=gray, ultra thick, rectangle, dotted, fill=yellow!01, minimum width=140, minimum height=15, align=left] (startthread) at (4.75,0.5) {\textbf{startThread(stmt, env)}\\\textbf{runTimeSlice(duration)}};
\node [draw=gray, ultra thick, rectangle, dotted, fill=yellow!01, minimum width=140, minimum height=15, align=left] (runtime) at (4.75,-0.75) {\textbf{runTimeSlideEnded()}};

\node [align=left] (orch) at (-2.1,2.4) {\textbf{Ozy Container}};

%\draw [dashed, ultra thick, ->] (in.south) -- (-2,2.15);
%\draw [dashed, ultra thick, ->] (out.south) -- (1,2.15);

\node [draw, circle, fill=blue!09, inner sep=6] (p) at (0.1,1.0) {P};
\node [draw, circle, fill=blue!09, inner sep=6] (tl) at (-1.3,-2.0) {Th};
\node [draw, circle, fill=blue!09, inner sep=6] (tc) at (1.5,-2) {Th};

\draw [ultra thick] (p) -- (0.1,1.7);
\draw [ultra thick] (0.1,1.7) -- (0.1,2.2);
\draw [ultra thick] (p) -- (tl);
\draw [ultra thick] (p) -- (tc);

\draw [ultra thick, dashed, ->] (0.4,1.5) -- (0.4,2.2);
\draw [ultra thick, dashed, ->] (-0.2,2.2) -- (-0.2,1.5);

\draw [gray] (createthread) -- (-0.2,2.2);
\draw [gray] (pterm) -- (0.4,1.5);

\draw [ultra thick, dashed, ->] (1.0,-0.1) -- (1.3,-0.8);
\draw [ultra thick, dashed, ->] (0.6,-1.0) -- (0.3,-0.3);

\draw [ultra thick, dashed, ->] (-0.8,-0.1) -- (-1.2,-0.8);
\draw [ultra thick, dashed, ->] (-0.4,-1.0) -- (-0.1,-0.3);

\draw [gray] (startthread) -- (1.0,-0.1);
\draw [gray] (runtime) -- (0.6,-1.0);

\node [draw=gray, thick, rectangle, fill=blue!10, minimum width=140, minimum height=25, align=left] (runtime) at (0.1,-4.35) {Dataflow Memory};
\node [draw=gray, thick, rectangle, fill=blue!10, inner sep=4, align=left, rotate=-90] (sr) at (0.5,-3) {Stack};
\node [draw=gray, thick, rectangle, fill=blue!10, inner sep=4, align=left, rotate=-90] (sl) at (-2.3,-3) {Stack};

\draw [] (tl.west) -- +(0,0)-| (sl);
\draw [] (tc.west) -- +(0,0)-| (sr);

\draw [very thick, <->] (tc.south) -- (1.5,-3.9);
\draw [very thick, <->] (tl.south) -- (-1.3,-3.9);

\draw [gray] (these) -- (tl);
\draw [gray] (these) -- (sl);
\draw [gray] (these.south) -- (runtime.west);

\draw [ultra thick] (-2.7, -5.5) -- (-1.9, -5.5);
\node [] (r) at (-0.9,-5.5) {Connector};

\draw [ultra thick, ->, dashed] (0.5, -5.5) -- (1.3, -5.5);
\node [] (r) at (2.2,-5.5) {Message};

\draw [very thick, <->] (-1.1, -6.0) -- (-0.3, -6.0);
\node [] (r) at (1.0,-6.0) {Private Access};

\end{tikzpicture}
    }
    \caption{Ozy Computation. If we zoom-in on Process and Thread from Fig.~\ref{fig:msgrouting}, we see the intersection of message routing and computation. Process and Thread implement the Oz Computation Model.}
    \label{fig:ozyorchestration}
\end{figure}

Figure~\ref{fig:ozyorchestration} shows the intersection of message routing and computation. After a message is correlated and translated into a semantic stack, semantic computation can begin. The first step of computation is performed when Ozy selects a time slice message for execution, thereby satisfying the OCM definition of thread selection: MST = \{ST\} $\uplus$ MST$^\prime$. The computation step is completed when a thread actor runs a time slice by popping its semantic stack and performing a semantic operation as a function of its environment and the single-assignment store (dataflow memory), thereby satisfying the OCM definition of a full computation step: (\{ST\} $\uplus$ MST$^\prime$, $\sigma$) $\rightarrow$ (\{ST$^\prime$\} $\uplus$ MST$^\prime$, $\sigma^\prime$).

\section{Case Study}

In this section we present two scenarios programmed with Ozy, which highlight the limitations of current approaches.

\begin{figure}[ht]
    %\centering
\hspace*{.25cm}
\begin{tikzpicture}
  \node(listing1) {
\begin{lstlisting}
proc {GetPrice Quantity ClientLoc Product ?Euro} SupId in
    {Orch.updateCSet cset(product:Product quantity:Quantity clientLoc:ClientLoc)}
    SupId = {Reg.getIdByQuery Product}
    (SupLoc1 SupLoc2) = {Reg.getData SupId}
    MyLoc = 'socket://localhost:2564'
    Euro = {Supp.getEuro SupLoc1 Product} * Quantity
    thread {Sleep 3000 millis} BuyTimeout=true end
    case {WaitTwo BuyOk BuyTimeout}
        of 1 then skip
        [] 2 then {Quantity ClientLoc Product 'No'}
    end
end
proc {Buy Quantity ClientLoc Product Conf} IdOrder BkId in
    BuyOk = true
    if Conf = 'Yes' then
        IdOrder = {Supp.order SupLoc2 Quantity ClientLoc Product}
        BkId = {Bank.pay Reg.bankLoc ClientLoc SupLoc2 MyLoc Euro}
        {Orch.updateCSet cset(bkId:BkId)}
        {Receipt BkId IdOrder}
        {Orch.commit ClientLoc}
    end
end
\end{lstlisting}
};
%\node[draw, ultra thick, fill=white, inner sep=.8em, rectangle, font=\large] (catavolt) 
\end{tikzpicture}

    \caption{Typical Scenario. A typical business process that establishes a correlation set, finds services in the registry, and coordinates interactions with other business processes.}
    \label{fig:typicalscenario}
    \vspace{-.5cm}
\end{figure}

Our first scenario in Fig.~\ref{fig:typicalscenario} comes from the original JOLIE paper~\cite{montesi2007composing}. This scenario is a typical business process involving five participants: a customer, a market, a service register, a supplier and a bank. Like JOLIE, this example demonstrates an approach more expressive and more programmer-friendly than using BPEL XML; however, unlike JOLIE, this example uses implicit synchronization on dataflow variables and pattern matching, which are not available in JOLIE, BPEL or other orchestration languages. 
Specifically, line 8 shows a pattern match implicitly synchronized on \texttt{BuyOk} and \texttt{BuyTimeout}. The pattern match waits on a nondeterministic choice from the \texttt{WaitTwo} procedure call. Computation is suspended at line 8 until either \texttt{BuyTimeout} is bound at line 7 or \texttt{BuyOk} is bound at line 14.

\begin{figure}[ht]
    %\centering
\hspace*{.25cm}
\begin{tikzpicture}
  \node(listing1) {
\begin{lstlisting}
proc {Subscribe ClientLoc TankId ?Stream} Tail in
    Stream = 0|Tail
    {Orch.updateCSet cset(clientLoc:ClientLoc
        tankId:TankId stream:Stream}
    {PushNextEvent TankId Stream}
end
proc {PushNextEvent TankId Stream} in
    thread {Sleep 1 minute} Proceed=true end
    case {WaitTwo Stop Proceed}
        of 1 then skip
        [] 2 then
            WaterLevel = {GetWaterLevel TankId}
            if {WaterLevelChanged WaterLevel 0.05} Next in
                case Stream of Head|Tail then
                    Tail = WaterLevel|Next end
            end
            {PushNextEvent TankId Stream}
    end
end
proc {Unsubscribe TankId Stream}
    Stop=true
end
\end{lstlisting}
};
%\node[draw, ultra thick, fill=white, inner sep=.8em, rectangle, font=\large] (catavolt) 
\end{tikzpicture}

    \caption{Non-typical Scenario. A non-typical Internet-of-Things scenario that emits an endless stream of integers to a waiting device until it unsubscribes.}
    \label{fig:stream}
    \vspace{-.5cm}
\end{figure}

Our second scenario in Fig.~\ref{fig:stream} is not a business process. Instead, it involves an Internet-of-Things device that monitors water level in a water tank. The service tracks the last water level reported and sends an update when the level changes by more than 5\%. Water levels are gathered on the local file system as comma-delimited text records. Line 2 initializes the \texttt{Stream} dataflow variable to a 0 water level value and an unbound tail. Line 3 establishes a correlation set with the \texttt{Stream} dataflow variable, which will be monitored by our connector. The procedure at Line 7 is called recursively (tail call optimized) from Line 17 to push water level events. Every time the \texttt{Stream} is extended, our connector is notified and transmits an integer to the device. This example demonstrates advanced programming features and flexibility not found in JOLIE, BPEL or other orchestration languages. Specifically, it uses a device connector based on a WebSocket where the connection is first established by the server for better security. Communication is a non-typical bi-directional connection that streams integers to the device only when needed at line 15, therefore avoiding inefficiencies and client-side polling. The service emits an endless stream of integers through the dataflow variable at line 15 until the connector unsubscribes at line 20.

\section{Conclusions}
Although BPEL XML and JOLIE have been used extensively to orchestrate business processes, they have major limitations that are succinctly yet powerfully addressed by applying lessons and technologies from the Oz computation model, most importantly implicit synchronization via dataflow variables and pattern matching. In this paper we present a description of the orchestration framework, Ozy, and provide its implementation as an open source project. We further show that the Ozy framework is easily used to program and perform standard business processes, and we also demonstrate the flexibility of Ozy with a long-running, implicitly synchronized, Internet-of-Things scenario.

This contribution opens the door for many possible follow on studies. One avenue for future work is in mobile devices and the Internet-of-Things, which are energy constrained. Ozy can be used for computational offloading, thereby reducing end-to-end latency between a device and its cloud services, which could be shown to improve the user experience and extend battery life.





\newcommand{\BIBdecl}{\setlength{\itemsep}{0.25 em}}
\bibliographystyle{IEEEtran}

%\bibliography{sigproc} 


\end{document}